**Title:** Large Language Models in Plant Biology
**Authors:** Hilbert Yuen In Lam[1], Xing Er Ong[1], Marek Mutwil[1]*

[1]School of Biological Sciences, Nanyang Technological University, 60 Nanyang Drive, Singapore, 637551, Singapore

*Corresponding author:
Marek Mutwil
School of Biological Sciences
Nanyang Technological University
60 Nanyang Drive
637551
Singapore

Email: mutwil@ntu.edu.sg

**Abstract:** Large Language Models (LLMs), such as ChatGPT, have taken the world by storm and have passed certain forms of the Turing test. However, LLMs are not limited to human language and analyze sequential data, such as DNA, protein, and gene expression. The resulting foundation models can be repurposed to identify the complex patterns within the data, resulting in powerful, multi-purpose prediction tools able to explain cellular systems. This review outlines the different types of LLMs and showcases their recent uses in biology. Since LLMs have not yet been embraced by the plant community, we also cover how these models can be deployed for the plant kingdom.

**Introduction**

Recent strides in deep learning have resulted in extraordinary feats in AI, such as providing accurate medical diagnostics, passing the bar exam, and completing certain versions of the Turing Test. The most well-known AI, ChatGPT, is a Large Language Model (LLM), a type of neural network able to generate text mirroring human language.  By training on billions of such texts, an LLM learns, through **self-supervised** means, a contextual understanding of the language it is trained on [1]. ChatGPT is based on a GPT (Generative Pre-trained Transformer) **foundation model** that was **pre-trained** on vast amounts of text data. LLMs are commonly pre-trained by solving cloze passages (e.g., "The ____ of France is Paris"), thus gaining an understanding of language and correlations between words. While foundation models are pre-trained to solve cloze passages, they can be **fine-tuned** to an array of purposes, allowing them to leverage previously learned knowledge on new problems. For example, ChatGPT is a fine-tuned version of a base GPT-3 model through **supervised learning**, where the model was trained on a dataset of conversational exchanges and is designed specifically for instruction-following tasks. Remarkably, ChatGPT was able to complement the "One hundred important questions facing plant science" list generated by a broad network of scientists, showing that LLMs can mimic knowledge and creativity [2,3].

LLMs can be adapted to analyze biological sequence data by treating DNA or amino acid sequences as text [4,5]. This has led to the popularizing of DNA and protein language models and other specialized LLMs for biological sequence data [6,7]. Similarly to LLMs applied to natural language, biological sequence LLMs are tasked to predict the identity of masked amino acids or nucleotides and consequently gain an understanding of the "protein/DNA language", allowing the models to find novel dependency patterns [8]. Similarly to the GPT-3 model, these biological foundation models can then be fine-tuned to a new task and show a remarkable aptitude to compete with and even surpass previous methods for protein structure prediction [4,5], gene function prediction [9,10], identification of regulatory elements and splice sites [7], protein design [11], and others [12,13] (**Figure 1**). This strong aptitude is attributable to the model applying knowledge gained through unsupervised learning into specific domains.

To understand biological systems, we often use simplified models and representations that are bound to result in digestive but incomplete narratives [14]. Conversely, LLMs excel at learning statistical properties of intricate, noisy sequential data and have the potential to model biomolecular systems that may even surpass their proficiency in modeling human language. With advances in genomic approaches that allow affordable and large-scale data generation, we find ourselves at the threshold of understanding the language of molecular biology.

However, LLM approaches have yet to be extensively applied in plant research. In this review, we will discuss the recent advances in using LLMs in biology and propose how the methods and available data could be used to bring new insights into plant research.

**What are Language Models?**

Language models (LMs) are algorithms or neural networks trained on large text datasets to learn the statistical patterns and relationships within natural language. They are a fundamental component of natural language processing (NLP) and have a wide range of applications in various fields, such as translation, text generation, and question-answering. While there is no clear definition of when an LM becomes an LLM, LLMs typically have more **parameters** (often billions), are built on larger **training data,** and possess more capabilities than LMs. LMs have been used for decades in biology, and they can be divided into word n-grams, convolutional neural networks (CNNs), Long Short-Term Memory (LSTM) networks, and transformers.

Word n-gram is a sequence of n consecutive words from a text, where n is a positive integer. For example, "cellulose synthase" is a 2-gram (or bi-gram). In biology, the word n-grams are typically used for text mining of scientific literature [15] and discovery of regulatory elements in DNA (where n-grams are k-mers are interchangeable)[16], protein-protein interactions [17], and others. However, these approaches disregard the word order and are thus unable to capture the context between the n-grams/k-mers.

Convolutional neural networks (CNNs) employ **convolutions**, which are kernels (filters) that are applied to images or sequences of characters to identify specific features or information. CNNs have been used in plant research to identify regulatory enhancers in DNA [18] and protein ubiquitination [19]. Nonetheless, like n-grams, CNNs are constrained by the size of the applied filters and are better at finding local patterns rather than long-range dependencies and complex sentence structures.

Long Short-Term Memory (LSTM) models are a type of recurrent neural network (RNNs) suitable for analyzing sequential data, such as text or biological sequences. LSTMs are able to capture long-range dependencies between more distant words in a sentence by using a combination of long and short-term memory constructs. LSTMs have been used in biology for annotating genomes and genotype classification [20]. However, since LSTMs and other RNNs take an output from a previous step as an input for the current step, they tend to "forget" the beginning of the text when text size increases due to the vanishing gradient phenomenon as the information gets compressed. LSTMs can also be unstable as they are affected by exploding gradients - making them particularly difficult to train for certain datasets [21]. Furthermore, since LSTMs process the words in a sequence, they are unable to take advantage of parallel computation, making their training slow and costly.

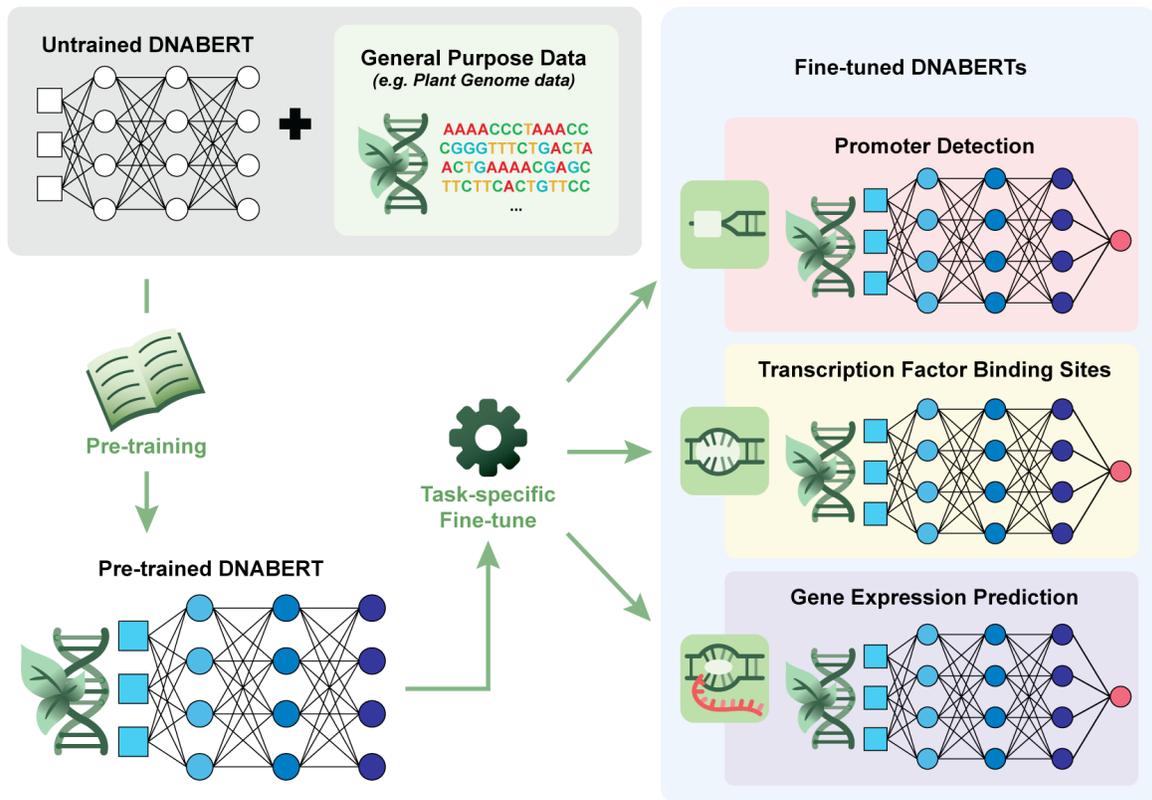

**Figure 1. Pre-training and fine-tuning of DNABERT model.** The model architecture, such as the number or layers and attention heads per layer, is initialized, and the model is pre-trained on sequential data, such as genomic DNA. The resulting pre-trained model can be repurposed to predict various DNA properties by fine-tuning.

Transformer models were introduced in 2017 to improve machine translation but have since been used to solve a variety of genomic approaches [6,7,22–24]. Transformer-based models typically outperform the abovementioned architectures by conferring several advantages. The main advantage in transformers is in their **multi-headed attention** mechanism. This is used in their **self-attention** mechanism which effectively captures long-range dependencies in the data, partially overcoming the "forgetting" problem of LSTMs, thus allowing longer sequences to be analyzed. Second, unlike RNNs, where computations depend on the previous step, transformers allow for parallel processing, making them more efficient for training, deploying, and upscaling [25]. Third, transformers allow the implementation of multi-headed attention mechanisms, where each head attends to a different part of the input text, allowing for a more nuanced understanding and better understanding of the long-range interactions [26]. Finally, these self-attention mechanisms can be **probed** to understand better which parts of the sequence a model focuses on, allowing the identification of the statistical relationships between elements in a sequence. However, the attention mechanism that underpins the high performance of transformers shows quadratic **complexity**, where the memory and computational requirements to calculate attention scale quadratically with the sequence length. This makes it computationally expensive to train transformers and effectively limits the length of sequences that can be analyzed.

**Examples of LLMs used in biology**
Using transformers in biology has allowed some recent breakthroughs, such as AlphaFold2. However, not all LLMs use transformers (e.g., HyenaDNA [27]), and not all models that use transformers are LLMs (e.g., AlphaFold2 [4]). BERT models are typically employed in classification, Named Entity Recognition (NER), and summarization, while GPT models are usually used for text generation and translation. However, fine-tuned versions of these two types can be used for tasks that are not originally intended, e.g., ChatGPT can be used for text classification and NER with **zero-shot** or **few-shot** prompting [28]. Since LLMs are used to analyze sequences of words, many types of sequential data can be used as training input, and in biology, both types of LLMs have recently been applied to study genomic, proteomic, and gene expression data. LLMs are typically pre-trained with self-unsupervised approaches that exploit the vast publicly available genomic data. BERT models employ Masked Language Modeling (MLM), where the aim is to predict masked **tokens**, while GPT models employ Causal Language Modeling (CLM), where the task is to predict the next token in a sequence. The predicted new word from the sequence can then be fed back into the same model iteratively to repetitively predict the next one - this makes a model **autoregressive**. Thus, pre-training aims to teach the model not to memorize the data but to learn to synthesize patterns that can be used to extract features and patterns and extrapolate to unseen data. The resulting foundation models can be repurposed for other tasks by fine-tuning them with supervised learning approaches [7,24,29]. The LLM models can be divided into three architecture types: encoder-decoder, encoder-only, and decoder-only (Box 1).

Encoder-only models
DNABERT is a BERT-family (Bidirectional Encoder Representations from Transformers) model trained with MLM, where the model is tasked with predicting the masked token by using the up- and down-stream tokens (hence bidirectional). During training, the DNA sequences are tokenized into k-mers, and the model is tasked to predict the masked k-mers by using the flanking k-mers (**Figure 1**). The training data can comprise species-specific genomes for species with multiple genomes (e.g., 1000 human genomes) or for multi-species genomes. DNABERT-2 was trained on genomes of 135 species totaling 32.49 billion nucleotide bases [23], while Nucleotide Transformer is trained on a staggering 3,202 human genomes and 850 genomes from other species [30]. Remarkably, these foundation models are multi-purpose tools that can be fine-tuned to excel at various tasks, such as identifying transcription factor binding sites, splice sites, and genes [7,23]. In addition, the models have been demonstrated to perform better on many tasks when trained on multiple diverse genomes compared to genomes trained on one species [7,23,30]. This indicates that LLM models based on different species can learn to capture genomic features of functional importance across species and, therefore, generalize better in various prediction tasks.

ESM-2 is a BERT-based (specifically RoBERTa [31], a derivative of BERT) LLM that is able to predict protein structures from single protein sequences accurately. ESM-2 was trained on ~65 million unique protein sequences with the MLM approach [32]. Remarkably, in contrast to AlphaFold2 (AF2) model that uses computationally expensive sequence alignments to identify pairs of correlated (and thus interacting) amino acids [33], ESM-2 model is simpler than AF2 and yet achieves a comparable performance while being up to 60x faster. The model was subsequently applied to predict structures of 772 million proteins from the kingdom of life (https://esmatlas.com/), allowing the study of protein structures on a metagenomic level.

The Geneformer model was trained on ~30 million human single-cell RNA-sequencing data, where gene expression values were transformed into sequences of gene IDs, by ordering the genes by their rank of expression within a cell [34]. The model was then tasked to predict masked genes in each cell by MLM and gained an understanding of the underlying gene regulatory network. After fine-tuning, the model gained state-of-the-art prediction performance on gene dosage sensitivity, chromatin dynamics, and network dynamics tasks.

Genomic Pre-trained Network (GPN) is the only model deployed specifically to plants [35]. The model architecture and training resemble DNABERT, but instead of transformers, it uses convolution mechanisms to capture the long-range dependencies between nucleotides. Like DNABERT, the model was able to identify DNA motifs, various types of genomic regions (intergenic, CDS, introns) and predict the effect of single-nucleotide polymorphisms. In contrast to popular tools that compute conservation scores (phyloP[36] and phastCons[37]), GPN can learn from joint nucleotide distributions across all similar contexts in the genome and does not rely on whole-genome alignments, which are often of lower quality in intergenic regions [35].

Decoder-only models

GPT family models have not been used much in biology. In addition to **self-attention**, decoders have a **cross-attention** mechanism. scGPT (single-cell GPT) is an LLM that achieves state-of-the-art performance across various downstream tasks [24]. To achieve this, the authors innovatively adapted the GPT technology by training the model on >10 million single-cell RNA-sequencing data. While GPT models excel in predicting the next word in a sentence, the authors tasked the model to predict the expression of genes in a cell, given a cell prompt and a subset of known genes expressed in the cell. The model achieved state-of-the-art performance on diverse tasks such as multi-batch integration, cell-type annotation, genetic perturbation prediction, and gene network inference by learning the intricate associations between cell types and gene expressions. Remarkably, the model can be readily extended to integrate multiple omic datasets (e.g., gene expression, chromatin accessibility, and protein abundance), working in an analogous mode as translating the "language" of one omic type to the 'language' of another.

**Model interpretability**

Model interpretability is a critical aspect of using Large Language Models (LLMs). **Interpretability** refers to the ability to understand and explain the decisions or predictions made by a machine learning model. In the context of LLMs, this is particularly important for several reasons. First, these models' high complexity and large number of parameters make them inherently difficult to understand. This "black-box" nature can be a significant hurdle in scientific settings where an understanding of how the model works is needed to gain insights into the underlying biological system [38]. Second, the ability to interpret models adds a layer of trust and reliability, essential for applications in sensitive areas like healthcare [39]. Third, interpretability can provide insights into the model's focus and decision-making process, which can be invaluable for fine-tuning and improving model performance.

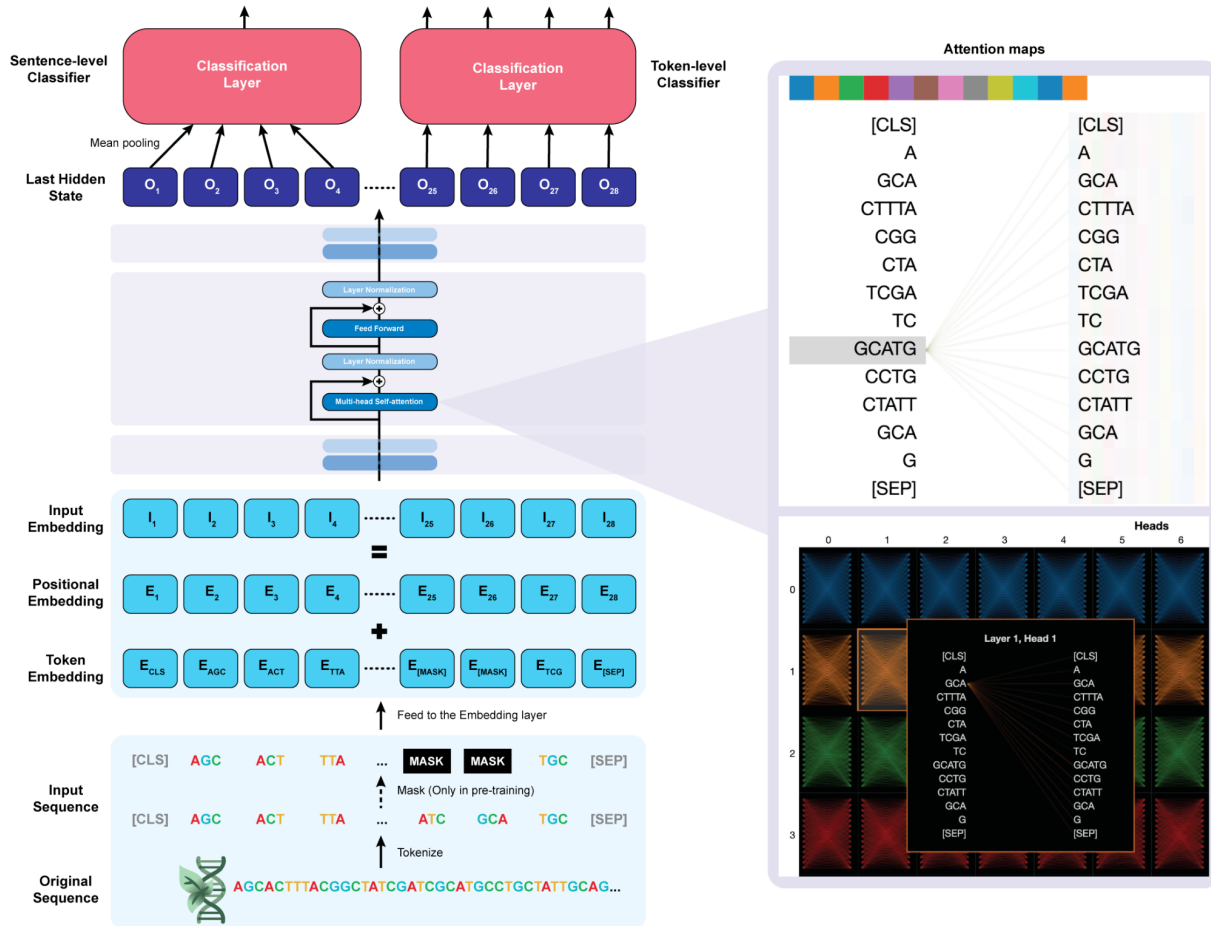

**Figure 2. Pre-training, anatomy, classification and probing of a DNABERT model.** The pre-training step accepts a DNA sequence that can be broken down into k-mers (tokenized), and the model is tasked with predicting the identity of masked tokens by Masked Language Modelling. The different tokens used in pre-training are: CLS token (describing the type of the particular sequence, e.g., CDS), a SEP token (separating the different sequences) and MASK tokens (masking k-mers in pre-training). The input passes an embedding layer and is processed by transformer blocks. The last hidden states can be used in a classification layer to predict properties of the analyzed sequence. DNABERT-2 attention maps visualized using BertViz enable better explainability of the model as to how certain tokens contribute to the contextual understanding of others.

Consequently, various techniques have been developed to improve the interpretability of LLMs [40,41]. These techniques include feature importance mapping, **attention maps**, and explainability frameworks that aim to break down the model's decisions into understandable components. For instance, the self-attention mechanisms in transformers can be probed to understand which parts of the sequence the model focuses on, thereby revealing statistical relationships within the data. In biological applications, this could mean identifying key genomic or proteomic features the model considers important for a particular prediction task (**Figure 2**).

Visualization tools also exist that can show the attention heads in BERT models such as BertViz [42], which can highlight which regions of the input sequence have the highest contribution in each layer of the model. In DNABERT [7], the authors developed DNABERT-viz, which likewise plots the self-attention maps of transformer heads. In scGPT, the authors likewise analyzed the attention maps, and also performed clustering, gene enrichment analysis on the embeddings as generated by the GPT model[24]. Through clustering and finding neighbors in the latent space, similarity networks could be constructed to infer interactions between discrete genes. This is a testament that although DNABERT and scGPT are two different lineages of transformer models, similar methods could be used to analyze them through unpooled embeddings or by deciphering the attention mechanisms.

In biological research, these properties of transformers could be used to in example (1) annotate DNA sequences and predict protein functionality, and then (2) explain which motifs in the sequence contribute to said prediction, hence providing insights into the key amino acid residues or nucleic acid motifs that give that property to the sequence. This could in turn provide leads for downstream wet lab validation experiments and improve overall understanding of these sequences.

**Building LLMs for plants - data, hardware, and software requirements**
Many of the aforementioned models have yet to be trained on plant data (e.g., DNABERT-2 was pre-trained on animals, bacteria, and fungi, but not plants), or have been only trained on a narrow selection of plant species. With more than 788 sequenced plant genomes [43], these models can be pre-trained on various plants spanning angiosperms, gymnosperms, ferns, lycophytes, bryophytes, and charophytes. Furthermore, as exemplified above, the LLMs can be pre-trained and fine-tuned with single-cell RNA-seq data (GeneFormer[34], scGPT[24]), with other modalities capturing epigenome [44], proteome [45] and metabolome [46] becoming more available. Just as Human Cell Atlas initiative that facilitates the annotation and interpretation of new datasets [47], recent efforts to collect and organize multimodal data will be invaluable in organizing the accumulating data for plants [48].

The available single-cell RNA-sequencing data for plants is accumulating, with more than 1 million sequenced nuclei available for *Arabidopsis thaliana*. These studies comprise atlases capturing seed-to-seed development (800,000 sequenced nuclei [49]), root brassinosteroid responses (210,856 nuclei [50]), root development (100,000 nuclei [51]), leaf infection time course (41,994 [52] and 11,895 [53] nuclei) and leaf UV treatment (23,729 nuclei [54]). Unsurprisingly, more data translates to a higher performance of LLM models, where scGPT LLM increased cell type annotation accuracy from 0.755 (300,000 sequenced nuclei) to 0.84 (3,000,000 nuclei [24]). Fortunately, the LLMs can account for batch effects, making it possible to integrate single-cell data from multiple studies [24,34]. This indicates that the already produced and future data will be invaluable to constructing plant LLMs.

The hardware requirements for LLMs are typically prohibitive, as these models require expensive GPUs with large amounts of memory [24]. Indeed, training models such as ESM-2 are not possible outside of large tech companies, where the training costs can exceed 200,000 USD (60 days on 512 NVIDIA® V100 GPU for the largest model [32]). However, computing can be affordable and easy to implement for smaller models and with the ubiquity of various cloud services. For example, pre-training DNABERT-2 would cost ~600 USD (14 days on 8 NVIDIA® GeForce® RTX 2080 Ti GPUs[23]), while GeneFormer would cost ~400 USD (3 days on 12 NVIDIA® V100 32GB GPUs[34]) on various cloud services (e.g, https://www.runpod.io/). Furthermore, most of the models are available as open-source software and are accompanied by user manuals, allowing easy pre-training and deployment for computational biologists fluent in Python (https://github.com/Zhihan1996/DNABERT_2, https://github.com/bowang-lab/scGPT).

**Concluding remarks and future perspectives**
The recent years have seen an emergence of powerful AI models that can sift through and identify patterns in noisy genomic data. LLMs have the potential to be divers of a paradigm shift, where the typically hypothesis-driven science can become increasingly data-driven. In this new paradigm, the researchers can start with a hypothesis-free, large-scale data generation that can be used to train an LLM.

LLMs have shown state-of-the-art performance in several predictive tasks, such as genome annotation, transcription factor binding site identification, and protein structure prediction. However, LLMs could provide deeper insights by integrating multi-modal data, offering a more holistic view of cellular systems and even stronger predictive performance. For example, by integrating chromatin accessibility and protein abundance measurements, scGPT performed better in identifying cell types [24]. At the same time, building models on data from multiple species can provide robust evolutionary insights, as exemplified by Nucleotide Transformer models' benchmarking experiments. The study showed that training with multiple human genomes did not perform as well as training with multi-species genomes, suggesting that multi-species models better capture functional importance conserved across evolution [30]. Finally, more data and increased diversity typically result in higher performance, as the predictive power of GeneFormer keeps increasing as a function of the number of cells in the training corpus [34].

The performance of the LLMs can also be increased by building models with more parameters, as larger models typically perform better [55]. However, since larger LLMs require more computing resources, this necessitates more efficient models. Fortunately, research into LLMs has brought several innovations, resulting in smaller, more efficient models. For example, FlashAttention uses an optimized read/write algorithm that speeds up the pre-training and allows for longer sequences to be analyzed [56], while sparse attention can significantly reduce the memory footprint and computational resources required for training [57]. For fine-tuning, Low-Rank Adaptation (LoRA) inserts small, trainable parameters into an otherwise pre-trained large model - reducing the memory footprint by up to three times [58]. More resource-efficient such as HyenaDNA forgo the expensive attention mechanism and use a long-dependency convolutional system, reducing training computation by 20% and increasing the number of tokens that can be interrogated to 1 million [27].

Given sufficient data and resources, LLMs have the potential to model cellular systems and extract insights about the underlying biological principles. Since cellular systems are marvelously complex, the LLMs could allow more accurate modeling of biomolecular systems at a granularity that goes well beyond human capacity [14]. We envision that more abundant data, efficient models, and increased adoption of LLMs will drive a data-driven paradigm shift in plant biology.

**Box 1. Different architectures of LLMs**
LLMs can be divided into encoder-decoder, encoder-only, and decoder-only architectures. The encoder-decoder architecture consists of two main parts: the **encoder** and the **decoder**. The encoder processes the input sequence and **embeds** it into a set of high-dimensional **latent space** vectors, capturing the meaning and context of the sequence. The decoder uses the latent space vectors to generate an output sequence, e.g., translation from one language to another. In transformer LLMs, the encoder and decoder have layers that include self-attention mechanisms, allowing them to consider the relationship between different words in the sequence. Thus, the encoder focuses on understanding the input, while the decoder focuses on generating the output. The different architectures excel at different tasks.

In the context of genomics, the Orca is an encoder-decoder able to predict chromosomal contact maps from DNA sequences. The Orca encoder accepts a one-dimensional DNA sequence and embeds it as a numerical vector, which is then decoded into a two-dimensional contact map that represents genomic proximity [59]. Encoder-only LLMs, such as BERT, excel at generating rich embeddings of biological sequences when the upstream and downstream context is important. These embeddings can be used as data for various classifiers. For example, DNABERT embeddings can be directly used to build state-of-the-art predictors of transcription factor binding sites, coding sequences, and mRNA splice sites [7]. Decoder-only architectures such as GPT excel at generating new sequences and various zero-shot prediction tasks. However, they can also be used for prediction tasks with fine-tuning, as exemplified by the scGPT (single-cell GPT) model that was trained on 33 million human cells and excelled at cell type annotation, genetic perturbation prediction, batch correction, and multi-omic integration [24].

**Highlights**
- Biological systems are incredibly complex and require novel approaches to be elucidated
- Large Language Models (LLMs) can find patterns and correlations in noisy biological data, but LLMs have not yet been embraced by the plant community
- We describe the different types of LLMs and cover how they can be used to study biological systems
- The data, hardware, and software requirements are within reach to build powerful LLMs for the plant community

**Outstanding Questions Box**
- How do we provide access to the accumulating multimodal single-cell data required to build LLMs?
- How do we lower the adoption thresholds for LLMs, making them easier to understand and deploy by the plant community?
- How much data and which data modalities are most needed to model cellular systems?
- Which insights can be gained by training LLMs on multimodal data from multiple species?

**Glossary**

**Self-supervised learning** uses data without explicit labels, for example by predicting masked tokens in DNA sequences.

**Foundation models** are general-purpose AI models trained on broad and diverse data. They can be **fine-tuned** for specific tasks.

**Fine-tuning** is the process of further training a **pre-trained** model on a more specific subset of data, to build e.g, a sequence classifier.

**Supervised learning** uses labeled data to train a model that can predict a desired property of an input.

**Pre-training** is used to create **foundation models** by **self-supervised learning**.

**Parameters** are typically trainable numerical variables inside AI models that determine how the model responds to input data.

**Training data** is used to train models and typically comprises DNA and protein sequences or gene expression.

**Multi-headed attention** builds upon the concept of self-attention by introducing parallel attention layers, or "heads" with each head focusing on different parts of the sequence, allowing the model to comprehensively understand the context of the entire sequence and be executed in parallel.

**Self-attention** is a process used in the **encoder** and **decoder** layers of **transformers**. It allows each element in the sequence (e.g., a word) to "attend" to other parts of the same sequence (other words in a sentence). This allows the model to understand the context of its input sequence.

**Cross-attention** is similar to **self-attention**, except the input sequence "attends" to parts of another sequence. It is only typically found in **decoders**, and allows the model to understand the context of sequence B given a query from sequence A.

**Probing** attempts to understand the prediction process of an LLM by analyzing the internal **embeddings** of the model.

**Complexity** describes how much computational resources are required to train or run a model as a function of its input size. **Transformers** have quadratic complexity due to the **multi-headed attention** mechanism.

**Autoregressive models** iteratively generate an output that is then fed back into the model as an input. In **causal language modeling**, it allows sequences to be continuously generated.

**Zero-shot, one-shot, few-shot, and many-shot learning** is when a model is trained with none, single, few, or many examples per class respectively.

**Tokens** are numerical representations of words, subwords or k-mers that can be used as input to LLMs.

**Convolutions** are kernels or filters that in the context of convolutional neural networks and genomics, can be trained to recognise certain DNA sequences.

**Interpretability** refers to the ability to explain the decisions made by a machine learning model.

**Attention maps** are visual representations of where a trained model is focusing or paying attention to input data. They can be used to determine which parts of an input are more "important" in determining its final output.

**Embeddings** are numerical values generated by an AI model used to represent an input, such as a sequence. They can be used to make downstream predictions in **probing**.